\documentclass[usenatbib]{mn2e}
\usepackage{apjfonts}
\usepackage{epsfig}
\usepackage{amssymb}
\usepackage{amsmath}
\usepackage{fixltx2e}

\newcommand\plotone[1]
 {\centering \leavevmode \includegraphics[width={0.99\columnwidth}]{#1}}
\newcommand{\plotside}[1]
 {\centering \leavevmode \includegraphics[width={0.95\textwidth}]{#1}}

\newcommand{\acknowledgments}{\begin{small}\section*{Acknowledgments}\end{small}}
\newcommand\altaffilmark[1]{$^{#1}$}
\newcommand\altaffiltext[1]{$^{#1}$}
\newcommand{\etal}{et al.}

\newcommand{\msun}{M_{\sun}}

\title[Spheroid Size Evolution]{Compact High-Redshift Galaxies Are the Cores of 
the Most Massive Present-Day Spheroids}

\author[Hopkins \etal]{
\parbox[t]{\textwidth}{ 
Philip F.\ Hopkins\altaffilmark{1}\thanks{E-mail:phopkins@astro.berkeley.edu},
Kevin Bundy\altaffilmark{1},
Norman Murray\altaffilmark{2,3},
Eliot Quataert\altaffilmark{1}, 
Tod R.\ Lauer\altaffilmark{4},
Chung-Pei Ma\altaffilmark{1}\
} 
\vspace*{6pt} \\
\altaffiltext{1}{Department of Astronomy, University of California 
Berkeley, Berkeley, CA 94720} \\
\altaffiltext{2}{Canadian Institute for Theoretical Astrophysics, 
60 St.\ George Street, University of Toronto, ON M5S 3H8, Canada} \\
\altaffiltext{3}{Canada Research Chair in Astrophysics} \\
\altaffiltext{4}{National Optical Astronomy Observatory, Tucson, AZ 85726}
}

\date{Submitted to MNRAS, March 3, 2009}
\begin{document}
\maketitle
\label{firstpage}

\begin{abstract}

Observations suggest that the effective radii of high-redshift massive spheroids 
are as much as a factor $\sim6$ smaller than low-redshift 
galaxies of comparable mass. Given the 
apparent absence of low-redshift counterparts, this has often been interpreted as 
indicating that the high density, compact red galaxies must be ``puffed up'' 
by some mechanism. We compare the ensemble of high-redshift observations 
with large samples of well-observed low-redshift ellipticals. 
At the same physical radii, the stellar surface mass densities of low and 
high-redshift systems are comparable. Moreover, the abundance of high 
surface density material at low redshift is comparable to or larger than 
that observed at $z>1-2$, consistent with the continuous buildup of spheroids 
over this time. The entire population of compact, high-redshift red galaxies 
may be the progenitors of the high-density cores of present-day ellipticals, 
with no need for a decrease in stellar density from $z=2$ to $z=0$. 
The primary difference between low and high-redshift systems is thus the 
observed low-density material at large radii in low-redshift spheroids (rather 
than the high-density material in high-redshift spheroids). Such low-density 
material may either (1) assemble at $z<2$ or (2) be present, but not 
yet detected, at $z>2$. Mock observations of low-redshift massive 
systems suggests that the amount of low-density material at high redshifts 
is indeed significantly less than that at $z=0$. However, deeper observations 
will be important in constraining the exact amount (or 
lack thereof) and distribution of this material, and how it builds up with redshift. 
We show that, without deep 
observations, the full extent of such material even at low redshifts 
can be difficult to determine, in particular if the mass profile is not 
exactly a single Sersic profile. We discuss
the implications of our results for physical models of galaxy evolution. 
\end{abstract}


\section{Introduction}
\label{sec:intro}

Recent observations suggest that 
high-redshift spheroids may have significantly smaller effective 
radii than low-redshift analogues of the same mass 
\citep[e.g.][]{daddi05:drgs,
trujillo:compact.most.massive,trujillo:ell.size.evol.update,
zirm:drg.sizes,
toft:z2.sizes.vs.sfr,
vandokkum:z2.sizes,
franx:size.evol,
vanderwel:z1.compact.ell,
cimatti:highz.compact.gal.smgs,
buitrago:highz.size.evol}. 
The apparent differences are dramatic: 
the inferred effective radii are as much as a factor $\sim6$ smaller 
at fixed stellar mass in the most massive galaxies at $z=2$. 
Whatever 
process explains this apparent evolution must be particular to 
this class of galaxies: disk galaxies are not similarly compact 
at high redshift \citep{ravindranath:disk.size.evol,ferguson:disk.size.evol,
barden:disk.size.evol,somerville:disk.size.evol}.
As such, these observations represent a strong 
constraint on models of galaxy and bulge formation. 

Relative to the abundance of massive galaxies today, there are not a large 
number of compact systems at high redshift. However, even if 
just $\sim10\%$ survived intact to $z=0$, this would greatly exceed the observed 
number density of such systems in the local Universe \citep{trujillo:dense.gal.nearby}. 
In fact, at fixed stellar mass, 
ellipticals with older stellar populations appear to have the 
largest radii \citep{gallazzi06:ages,bernardi:bcg.scalings,graves:ssps.vs.fp.location,
vanderwel:size.numden.massive.ell}. 

The challenge for both observations and models is therefore to understand how 
these high-redshift systems could evolve to become ``typical'' spheroids 
today. Their masses, number densities, and clustering 
dictate that they are the progenitors of the most massive ellipticals and 
BCGs today \citep{quadri:highz.color.density,hopkins:clustering}. 
These systems have much larger $R_{e}$ and thus have 
lower {\em effective} densities $\Sigma_{\rm eff}=M_{\ast}(<R_{e})/(\pi\,R_{e}^{2})$. 
However, this does not necessarily mean that the {\em physical} densities are 
lower than those of the high-redshift systems. One way to increase $R_{e}$ would be to 
uniformly ``puff up'' the profiles, lowering the physical density everywhere.
This would imply that the central densities of massive high-redshift ellipticals 
would need to decrease 
by two orders of magnitude from $z=2$ to $z=0$. 
Alternatively, $R_{e}$ can change by just 
as much by adding a relatively small amount of mass at low surface densities 
and large radii, without affecting the central density at all. In other words, an evolving 
effective density does not necessarily imply an evolving physical density at all radii. 

Buildup of an ``envelope'' of low-density material is expected as 
massive early-forming 
galaxies undergo late-time (major and minor) gas-poor mergers with 
later-forming, less-dense ellipticals, disks, and dwarfs 
\citep{gallagherostriker72,ostrikertremaine75,hausman:mergers,
weil94:multiple.merger.kinematics,weil96:multiple.merger.scalings,naab:etg.formation}. 
This is a relatively efficient channel for size-mass evolution, yielding factor 
of several size evolution with only a factor $1.5-2$ increase in stellar mass 
\citep{hopkins:cores}. But this less dense material added at large radii does not 
significantly affect the high-density core,\footnote{By ``core,'' we refer to 
the central regions of the galaxy, not to any specific 
class of central profile slopes. We use the phrases ``cusp ellipticals'' or 
``core ellipticals'' to distinguish these.} and so these models predict that the dense, 
high-redshift systems should survive to become the central regions of 
(some fraction of) today's massive ellipticals. 

If, on the other hand, high-redshift systems evolved primarily by
equal-mass dry mergers between equivalently dense spheroids, then this will
``inflate'' the profiles relatively uniformly.  In this extreme case,
effective radii and stellar mass both approximately double in the merger;
high-redshift systems would be uniformly more dense than their low-redshift
descendants \citep[see e.g.][]{hernquist:phasespace,
boylankolchin:dry.mergers}.

Distinguishing between these possibilities, as well as other 
dynamical, stellar evolution, or 
observational effects that could lead to apparent size-mass evolution 
clearly depends on understanding in detail differences in the surface density 
profiles of spheroids as a function of radius, at low and high redshift. 
In this paper, we quantitatively compare low and high-redshift observations 
to constrain these scenarios and inform how systems are evolving 
from $z\gtrsim2$ to $z\sim0$. 

In \S~\ref{sec:prof.compare} we directly compare the observed profiles 
of high and low-redshift massive spheroids, and show that, at the same 
physical radii, their stellar surface mass densities are comparable. 
The massive, high redshift systems appear no different than the ``cores'' 
of today's massive ellipticals. 
In \S~\ref{sec:max.density} we determine the distribution of maximum/central 
densities, and show that this has not evolved significantly from 
$z=0$ to $z>2$. In \S~\ref{sec:rho.vs.sigma} we calculate the 
mass function (and global mass density) of these high-density ``cores'' 
at both low and high redshift. This allows us to quantitatively compare 
the abundance of 
high-density material observed at both low and high redshifts. 
We show that there is as much or more high-density material 
in the cores of massive spheroids at $z=0$ as is observed to be in place 
at $z=1-2$. The difference between low and high-redshift 
systems, we conclude, lies in the lack of observed low surface-density 
envelopes around the high-redshift systems. 
In \S~\ref{sec:obs.fx} we show that, although 
such envelopes are weaker at high redshifts, some caveats should 
be born in mind with respect to 
the determination of the mass in low-density material in 
high-redhsift spheroids. 
We summarize our results and discuss their consequences 
for physical models of spheroid evolution in \S~\ref{sec:discussion}. 

Throughout, we assume a WMAP5 cosmology \citep[$\Omega_{\rm M}=0.27$, 
$\Omega_{\Lambda}=0.73$, $h=0.705$;][]{komatsu:wmap5}, 
but the exact choice makes no significant difference.

\section{Surface Density Profiles of Ellipticals 
at High and Low Redshift}
\label{sec:prof.compare}

\begin{figure*}
    \plotside{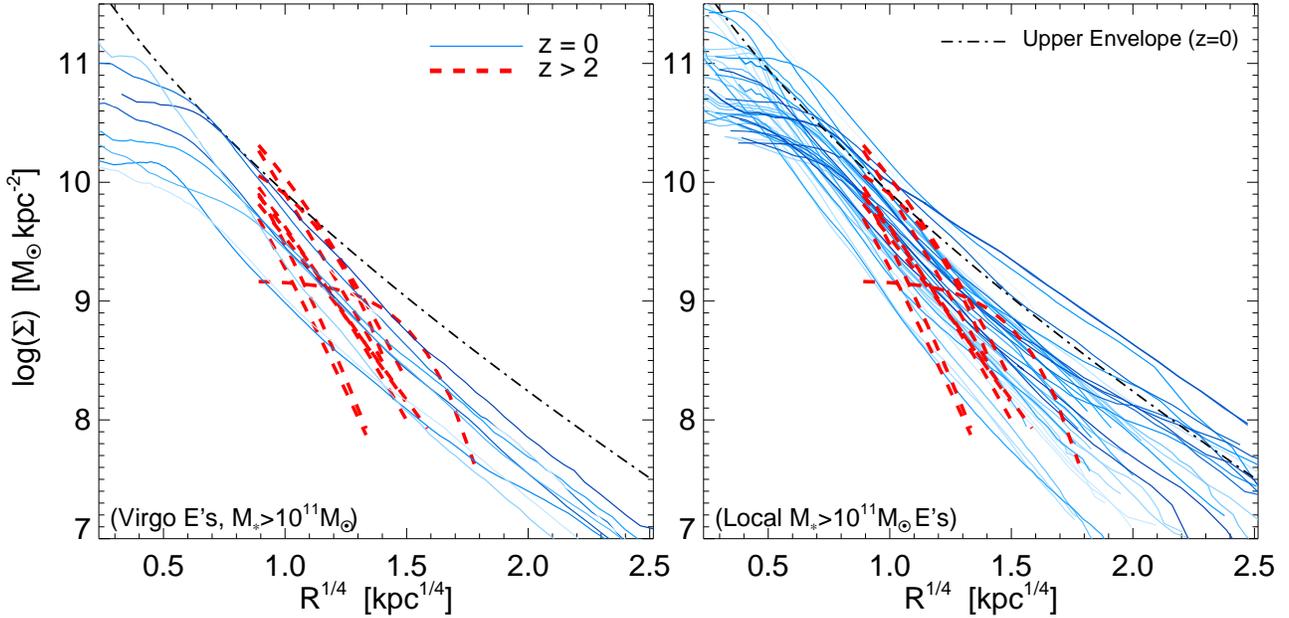}
    \caption{Direct comparison of the (major axis) surface stellar mass density profiles of 
    $z\gtrsim2$ compact massive spheroids ($M_{\ast}\sim10^{11}\,\msun$, 
    best-fit $R_{e}\sim$kpc; red dashed) and local, massive ellipticals 
    ($M_{\ast}\ge 10^{11}\,\msun$, $R_{e}\gtrsim4-5\,$kpc; solid).  
    Both $\Sigma$ and $R$ are in physical units.
    The high-$z$ profiles are the PSF de-convolved fit, plotted over the 
    observed range in radius given the best-case limitations in seeing and 
    surface brightness \citep[from][]{vandokkum:z2.sizes}.
    The low-$z$ profiles 
    combine space and ground-based photometry to obtain very large 
    dynamic range; they are from 
    the \citet{jk:profiles} Virgo elliptical sample ({\em left}) and the 
    \citet{lauer:bimodal.profiles} local massive elliptical sample ({\em right}). The former 
    is a volume-limited sample, so contains fewer very massive galaxies with high 
    central surface brightness. Although the high-$z$ systems have much smaller 
    $R_{e}$, their densities at any physical radius are not unusual 
    compared to the local objects: the central $\sim1-2$\,kpc
    of massive ellipticals today are just as dense. The difference in $R_{e}$ owes to 
    the presence of the large wings/envelopes at low surface density in the 
    low-redshift objects.
    \label{fig:profiles.compare}}
\end{figure*}

Figure~\ref{fig:profiles.compare} shows a direct comparison 
of the observed surface stellar density profiles of high-redshift 
compact galaxies and low-redshift massive galaxies. 
At low redshift we compile observed surface brightness profiles from 
\citet{jk:profiles} and \citet{lauer:bimodal.profiles}; this consists of 
a total of $\sim180$ unique local 
ellipticals with nuclear {\em HST} observations and 
ground-based data at large radii 
(allowing accurate surface brightness profile measurements from 
$\sim 10$\,pc to $\sim50$\,kpc).\footnote{Note 
that although the 
composite (HST+ground-based) profiles 
were used in \citet{lauer:bimodal.profiles} to estimate effective radii, 
they were not 
actually shown in the paper.} 
The isophotally averaged major axis profiles are measured in rest-frame 
optical; we convert to a stellar mass profile based on the measured 
total stellar masses and the assumption of a radius-independent 
stellar mass-to-light ratio. Conversion to stellar 
mass profiles using e.g.\ color or stellar population gradients 
and comparison of profiles from different 
instruments and wavebands in these samples are discussed extensively 
in \citet{hopkins:cusps.ell,hopkins:cusps.fp}; the differences 
are much smaller than the scatter between individual profiles, and 
do not affect our conclusions. 
In Figure~\ref{fig:profiles.compare}, 
we restrict our comparison to massive 
galaxies with $M_{\ast}>10^{11}\,\msun$\footnote{Stellar masses for all 
objects are determined from the combination of rest-frame optical and 
near-IR photometry, corrected to an assumed \citet{chabrier:imf} IMF. 
We refer to \citet{hopkins:cusps.ell} and \citet{kriek:08.nir.spectroscopy.highz} 
for details of the low and high-redshift samples, respectively. 
Varying the specific bands used to determine stellar masses makes little 
difference, and changing the IMF will systematically change the stellar 
masses of all objects considered, but will not change our comparisons.}
because these systems are 
most likely to be descendants of massive high-redshift galaxies. 
The \citet{jk:profiles} sample is a volume-limited survey of the 
Virgo spheroid population; as such it includes few very massive 
galaxies ($M_{\ast}>3\times10^{11}\,\msun$). The \citet{lauer:bimodal.profiles} 
galaxies are chosen to be representative of massive ellipticals 
in the local Universe, including more massive systems up to 
a couple $10^{12}\,\msun$. At the masses of interest, both are 
representative of the distribution of spheroid sizes in the local SDSS galaxy 
sample \citep{shen:size.mass}. 

Figure~\ref{fig:profiles.compare} 
compares the low-redshift sample with the observed, PSF 
de-convolved profiles of nine
high-redshift compact massive galaxies ($M_{\ast}\gtrsim10^{11}\,\msun$, 
$R_{e}\sim1\,$kpc), specifically the $z\sim2-3$ sample 
from \citet{vandokkum:z2.sizes}. 
This is a well-studied sample that represents 
the extreme of implied size evolution: the inferred average $R_{e}$ is a 
factor of $\sim6$ smaller than local spheroids of the same mass. 
Figure~\ref{fig:profiles.compare} shows 
the best-fit Sersic profile of each galaxy in the sample; stellar mass-to-light 
ratios are determined by assuming a radius-independent 
$M_{\ast}/L$ and normalizing the observed portion of the profile to the total 
stellar mass determined from photometry and spectroscopy in 
\citet{kriek:drg.seds,kriek:08.nir.spectroscopy.highz,kriek:highz.red.sequence}. 
We plot the profile of each system over the maximum radial range observed: 
from the scale of a single pixel at the observed redshift to 
the limiting surface brightness depth of the best images. Both the 
low and high-redshift systems are plotted in terms of major-axis radii 
(a non-negligible correction). 

At low redshift, the stellar mass-to-light ratios of ellipticals appear to
be nearly independent of radius (reflected in e.g.\ their observed weak
color gradients), but the stellar mass-to-light ratio may depend
significantly on radius in the high-redshift systems
\citep{trager:ages,cote:virgo,sanchezblazquez:ssp.gradients}.  However,
based on the observed stellar population gradients in local ellipticals,
the observed ages/colors of the high-redshift systems, or the outcomes of 
numerical simulations, the expected variation in $M_{\ast}/L$ is such that
a young, recently-formed post-starburst stellar population at the center of
the high-redshift galaxy will have higher $L/M_{\ast}$ than older stars at
larger $R_{e}$ \citep[see e.g.][]{hopkins:cusps.mergers}.  This would make
the high-redshift systems less dense than we assume here; we conservatively
allow for the maximal stellar mass densities in those systems.

The comparison in Figure~\ref{fig:profiles.compare} is quite striking: 
although the best-fit effective radii and 
effective surface densities of the high-redshift systems are quite different from 
their low-redshift analogues, the actual stellar surface mass densities at any given 
observed radius do not appear significantly higher than a substantial 
fraction of the low-redshift population. In other words, inside the same observed radii 
$\sim1-5$\,kpc, many of today's massive ellipticals are just as dense as 
the high-redshift systems. The difference in effective radius stems primarily from 
the fact that the low-redshift systems have substantial extended wings/envelopes of 
low surface-brightness material ($\Sigma\ll 10^{9}\,\msun\,{\rm kpc^{-2}}$); 
by contrast, the inference from fitting the high-redshift systems 
is that their profiles fall more rapidly at large radii (as we discuss further 
below). 

For the sake of comparison with these and future high-redshift observations, it 
is useful to define the ``upper envelope'' of low-redshift galaxy density profiles. 
Formally, we can take e.g.\ the $+1\,\sigma \approx 86\%$ contour of the combined 
sample of density profiles shown in Figure~\ref{fig:profiles.compare}, but we 
find that this can be conveniently approximated with a simple Sersic function. 
This envelope is approximately given by 
\begin{equation}
\Sigma_{+1\,\sigma}(z=0) \approx  4.5\times10^{12}\,M_{\sun}\,{\rm kpc^{-2}}\,
e^{-11.67\,{\bigl(} \frac{R}{40\,{\rm kpc}} {\bigr)}^{1/6}}\ . 
\label{eqn:do.not.exceed}
\end{equation}
This is a Sersic profile with $n_{s}=6$, $R_{e}=40\,{\rm kpc}$, 
and total stellar mass $M_{\ast}=1.7\times10^{12}\,\msun$. 
Note that most {\em single} galaxies do not remain 
along this envelope over its entire extent. 
Rather, at each radius, this represents the $+1\,\sigma$ upper extent of 
observed densities within the spheroid population (i.e.\ the most dense systems at each radius); 
Most individual systems approach it over some more limited dynamic range. 
This is a useful comparison quantity in particular because, although the {\em mean} profiles of 
$z=0$ ellipticals are well-known, there has been relatively little parameterization of the 
scatter in profile shapes. Thus, even if high-redshift ellipticals are more dense than 
the median system today, if they do not exceed the relation given by Equation~\ref{eqn:do.not.exceed}, 
then their densities can be accommodated within some portion of the present-day 
spheroid population, so long as they do 
not represent a large fraction of the present-day abundance of spheroids. 
As expected, we find that at their centers, the $z=2$ systems are at most comparable to this 
upper envelope, i.e.\ comparable to the most dense $z=0$ spheroid cores, 
and at large radii, they fall well below the envelope.

\section{Central Stellar Densities}
\label{sec:max.density}

\begin{figure*}
    \plotside{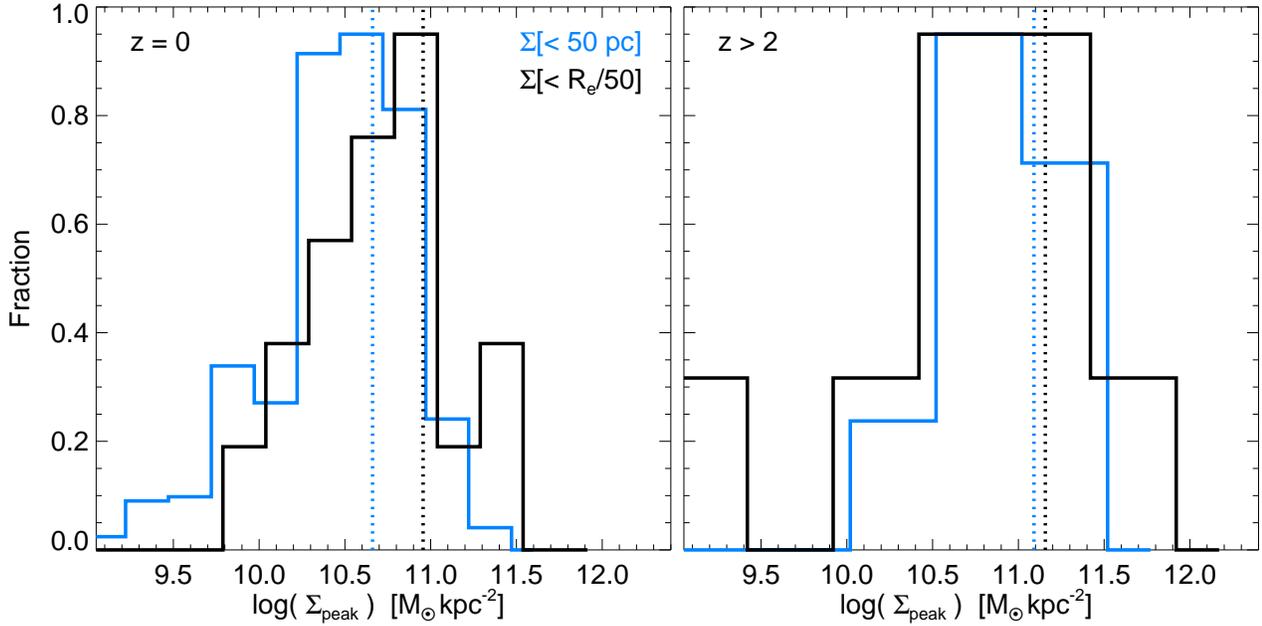}
    \caption{Distribution of peak (maximum) spheroid stellar mass 
    surface densities. {\em Left:} Low-$z$ 
    systems from Figure~\ref{fig:profiles.compare}. Most ellipticals, 
    at any mass, have $\Sigma_{\rm peak}\sim0.3-1\times10^{11}\,\msun\,{\rm kpc^{-2}}$. 
    Histograms show two 
    different calculations of $\Sigma_{\rm peak}$ (within 
    $50\,$pc and $R_{e}/50$); dotted lines are the median from each. 
    {\em Right:} Same, for the high-$z$ samples. 
    Direct observations are PSF and seeing-limited, so we extrapolate the best-fit 
    Sersic profile inwards to obtain $\Sigma_{\rm peak}$ (this is typically an upper limit 
    in the low-$z$ samples). 
    The results are striking: central densities are reasonably independent of redshift. 
    \label{fig:peak.densities}}
\end{figure*}

Figure~\ref{fig:peak.densities} plots the {\em peak} 
surface stellar mass densities $\Sigma_{\rm peak}$ obtained in both 
the low and high-redshift galaxy populations. 
The ``maximum'' or peak surface density must be defined within 
some radius, for galaxies whose surface density continues to 
rise to unresolved radii (e.g.\ cusp ellipticals; although most of the 
local massive galaxies are core ellipticals, with relatively 
flat maximum surface densities within $\sim50-500\,$pc). 
For example, the maximum stellar surface density can be defined 
as the average $\Sigma$ interior to some fixed 
small radius $\sim50-100\,$pc, or a fixed fraction of $R_{e}$ 
from $\sim0.02-0.04$, or by extrapolation to $R=0$ of a 
best-fit Sersic profile -- we are simply interested in 
comparing the central densities at small radii in both low and high-redshift systems. 
For each determination, we find qualitatively similar 
results albeit with some small normalization differences; Figure~\ref{fig:peak.densities} 
shows $\Sigma_{\rm peak}$ determined from averaging within 
$50\,$pc and within $0.02\,R_{e}$. For the high-redshift systems, 
PSF and seeing effects smear out the 
maximum observed surface brightness/density inside 
$\sim 1\,$kpc; we extrapolate the best-fit 
Sersic profiles inwards to the same radii as the local systems. 
Making this same approximation in the low-redshift samples shows that it is 
reasonable, but tends to slightly 
over-estimate the central surface density, especially in core ellipticals.  

At both low and high redshift, there is a characteristic maximum 
central surface stellar mass density $\sim0.3-2\times10^{11}\,\msun\,{\rm kpc^{-2}}$, 
with significant, but still surprisingly little scatter given the known diversity in 
the profile shapes of ellipticals (e.g.\ 
variation in Sersic indices and cusp versus core 
populations). The maximum is similar whether we include ellipticals of all masses 
($\sim10^{9}-10^{12}\,\msun$), or restrict to cusp or core populations; 
although dry mergers are expected to transform cusp ellipticals into 
core ellipticals via ``scouring'' (ejecting mass from the central regions in a 
BH-BH merger), this primarily affects the mass profile on very small 
scales (comparable to or less than the scales here -- well below what 
we generally refer to as the ``central'' regions of ellipticals at $\sim$kpc 
scales). 
The maximum surface density of the high-redshift systems is 
perhaps a factor $\sim2$ larger than that of low-redshift 
systems, but given that we are extrapolating Sersic profiles 
inwards for the high-redshift systems, this is probably an upper limit. Thus while 
the observations imply up to a factor $\sim40$ evolution in effective 
surface brightness, 
there is nowhere near this much evolution in the true maximum stellar 
surface density, a much more physically relevant quantity. 

Recently, \citet{bezanson:massive.gal.cores.evol} reached very similar conclusions 
from a similar comparison 
between high and low-redshift profiles, using a 
different methodology and low-redshift sample. One subtle 
difference between our conclusions and theirs is that they claim 
the central densities of the high redshift systems appear higher, on average, 
than those at low redshifts by a small factor ($\sim$ a couple). 
This owes largely to the fact that the authors extrapolate the best-fit 
Sersic profiles inwards to small radii -- as we show here, this does lead to 
slightly higher central densities. But we caution that, doing the same 
in the low-redshift systems, we would obtain similar slightly higher 
densities: over the observed range (Figure~\ref{fig:profiles.compare}) 
there is no difference. Moreover, \citet{bezanson:massive.gal.cores.evol} compare 
the high-redshift profiles only with the average profile of similar-mass 
galaxies at low redshift; in Figure~\ref{fig:profiles.compare} and 
\S~\ref{sec:rho.vs.sigma} we consider the full distribution of 
profile shapes at low redshift: even if the high-redshift systems 
have, on average, slightly higher central densities, they are still 
compatible with the central densities of a large fraction of the 
$z=0$ massive elliptical population. 

It is worth noting that the mass and redshift-independence in $\Sigma_{\rm peak}$ in
Figure~\ref{fig:peak.densities} is somewhat surprising, 
given the diversity of formation histories and scatter
in e.g.\ the mass present at larger radii \citep[see
e.g.][]{hopkins:msigma.scatter}.  A more detailed discussion of this will
be the subject of future work, but it may relate to the maximum surface
density of gas that can turn into stars (see \S~\ref{sec:discussion}).

\section{The Mass At High Stellar Densities}
\label{sec:rho.vs.sigma}

\begin{figure}
    \plotone{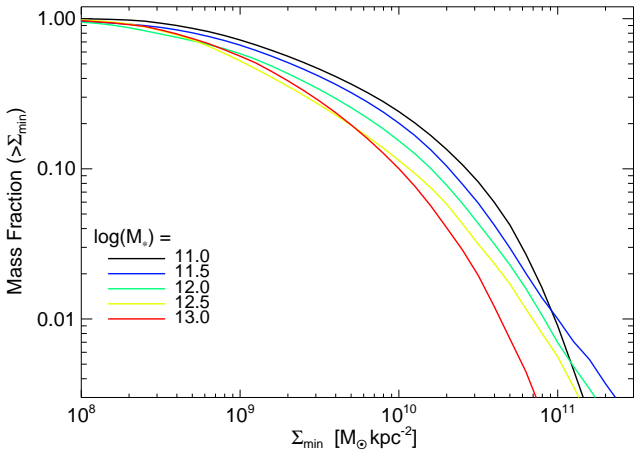}
    \caption{Average mass fraction above some threshold 
    surface stellar mass density $\Sigma_{\rm min}$, 
    for the low-$z$ sample from Figure~\ref{fig:profiles.compare}. The 
    values shown are the average fraction of {\em all} spheroid mass 
    (in a narrow range 
    around each $M_{\ast}$) above the given $\Sigma_{\rm min}$. A significant 
    fraction of the $z=0$ spheroid mass ($\sim25\%$) resides in matter at 
    the inferred effective $\Sigma$ of the high-$z$ compact systems 
    ($10^{10}\,\msun\,{\rm kpc^{-2}}$). 
    \label{fig:frac.above.sigma}}
\end{figure}

We now quantify the amount of mass at different stellar surface densities. 
In order to reduce the effects of noise and PSF effects 
(important in particular for the high-redshift 
systems), we define the surface density in this section 
as the average surface density within each radius, 
i.e.\ $\Sigma(R)=\langle\Sigma(<R)\rangle = M_{\ast}(<R)/\pi R^{2}$. 
We obtain similar results using the local $\Sigma$, 
but with larger noise. 
For each observed system in our low-redshift sample, given the stellar mass profile 
(Figure~\ref{fig:profiles.compare}), we calculate the 
total fraction of the stellar 
mass that lies above a given threshold in surface density 
$\Sigma_{\rm min}$. 
We evaluate this for each system separately, and in Figure~\ref{fig:frac.above.sigma} 
we plot the average mass fraction at each $\Sigma$ for all observed systems in our sample, 
in several bins of total stellar mass. Although the mass fraction above 
each $\Sigma_{\rm min}$ can 
vary by a large amount from galaxy to galaxy, the 
average is surprisingly robust across masses (and does not depend significantly 
on whether we include both cusp and core ellipticals or evaluate the two 
separately). By sufficiently low $\Sigma_{\rm min}$ thresholds, $\sim10^{8}\,\msun\,{\rm kpc^{-2}}$, 
essentially all mass in spheroids is accounted for; some 
systems have more extended, 
lower surface-brightness envelopes, but they contribute little total mass. 
Approximately $\sim25\%$ of the stellar mass density at each mass 
remains above $\sim10^{10}\,\msun\,{\rm kpc^{-2}}$ -- a typical 
effective surface brightness for high-redshift ellipticals -- in moderately 
high-mass systems (dropping to $\sim10\%$ 
by the most massive $10^{13}\,\msun$ systems). 
By a threshold of $\sim10^{11}\,\msun\,{\rm kpc^{-2}}$, 
we have reached the maximum/peak surface 
densities of ellipticals (Figure~\ref{fig:peak.densities}), 
and the mass fractions at higher densities drop rapidly. 

The best-fit Sersic profiles of the $z\sim2$ systems imply that 
they have higher mass {\em fractions} 
above a high surface density threshold $\sim10^{10}\,\msun\,{\rm kpc^{-2}}$. 
However, as illustrated above, this primarily owes to their having less 
mass at low $\Sigma$, not more at high $\Sigma$. 

\begin{figure}
    \plotone{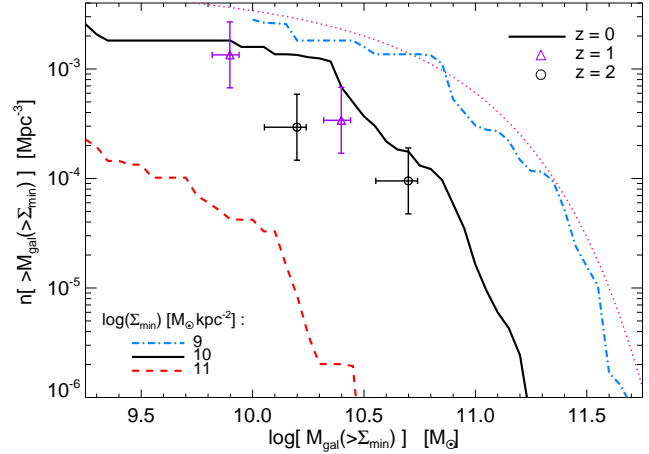}
    \caption{Spheroid stellar mass function (number density) for mass above 
    a given surface stellar mass density threshold ($\Sigma_{\rm min}$).
    We calculate the $z=0$ mass function of spheroids including 
    {\em only} stellar material in any galaxy above 
    the given $\Sigma_{\rm min}$, given the total stellar mass 
    function and the observed 
    distribution of profile shapes. 
    The magenta dotted line shows the total spheroid mass function 
    \citep{bell:mfs} -- most of the 
    mass is accounted for above $10^{9}\,\msun\,{\rm kpc^{-2}}$. 
    The value of $\Sigma_{\rm min}=10^{10}\,\msun\,{\rm kpc^{-2}}$ corresponds 
    to the effective 
    surface density of the compact high-redshift spheroids; points show 
    the results of the same calculation for the observed high-redshift systems 
    and this value of $\Sigma_{\rm min}$. We show this for the $z\sim2.3$ data 
    from \citet[][black circles]{vandokkum:z2.sizes} and 
    $z\sim1$ data from \citet[][triangles]{vanderwel:z1.compact.ell}. 
    \label{fig:mf.versus.sigma}}
\end{figure}

The comparison between the low and high-redshift samples can be 
made more quantitative by determining the 
stellar mass function above a given surface density threshold. To do so, we
ignore all stellar mass in the Universe below a given threshold in 
$\Sigma$, and construct the spheroid mass function. The mass of a 
given galaxy is only the mass above that $\Sigma$; 
i.e.\ we calculate the volumetric number density of spheroids 
with $\Sigma>\Sigma_{\rm min}$ 
\begin{equation}
n[>M_{\rm gal}(>\Sigma_{\rm min})] \equiv \frac{{\rm d}N({\rm galaxies} | M_{i}
> M_{\rm gal})}{{\rm d}V}
\end{equation}
as a function of the integrated 
mass above $\Sigma_{\rm min}$, 
\begin{equation}
M_{i}\equiv M(>\Sigma_{\rm min}) = 
\int_{\Sigma=\Sigma_{\rm min}}^{\Sigma\rightarrow\infty} 
\Sigma\times2\pi\,r\,{\rm d}r.\ 
\end{equation}
The resulting mass 
functions are shown in Figure~\ref{fig:mf.versus.sigma}. 
In detail, we take the observed stellar mass function of 
spheroids \citep[][]{bell:mfs}, and at each mass, convolve with the distribution of 
surface density profiles from \citet{jk:profiles} and \citet{lauer:bimodal.profiles} 
for systems of the same mass, to determine the resulting mass function (number 
density) 
above a given surface density threshold.\footnote{\label{footnote:densitydef}
Note that ``density'' here has two meanings: the volume density (the 
$y$-axis of Figure~\ref{fig:mf.versus.sigma}) is the total 
number of galaxies meeting a given criteria per unit volume; 
the surface density 
is the local stellar surface mass density of stars within a 
particular galaxy.}
We are assuming that 
the distribution of profile shapes in \citet{jk:profiles} and \citet{lauer:bimodal.profiles} 
are representative at each mass. Their sample 
selection as well as other measurements 
\citep[see those works and e.g.][]{trujillo:sersic.fits,ferrarese:profiles,
allen:bulge-disk} and the 
close agreement in fundamental plane correlations and 
Sersic index distributions in larger volumes \citep[e.g.][]{shen:size.mass} 
suggest that this is probably a good assumption. 

Unsurprisingly, Figure~\ref{fig:mf.versus.sigma}
shows that at lower $\Sigma$ thresholds the predicted mass 
function converges to the total stellar mass function of spheroids, i.e.\ most 
stellar mass is accounted for. And at high thresholds it drops rapidly, especially 
at high masses: $10^{11}\,\msun\,{\rm kpc^{-2}}$ is the peak surface density 
inside $\ll 1\,$kpc at both low and high redshifts -- no observed  
systems have significant amounts of 
mass $>10^{11}\,\msun$ above this threshold. 

In Figure~\ref{fig:mf.versus.sigma} we compare the 
$z=0$ volume density of mass at $\Sigma>10^{10}\,\msun\,{\rm kpc^{-2}}$ 
(solid lines) with observational inferences at $z=1$ (diamonds) 
and $z=2$ (circles). Specifically, 
given the {\em total} number density  
of $>10^{10}\,\msun$ spheroids/compact red systems 
at these redshifts \citep{perezgonzalez:mf.compilation}, 
convolved with the distribution of profile 
shapes from \citet{vandokkum:z2.sizes}, we obtain the number density of 
objects with mass above the relevant $\Sigma$ threshold. 
In the error budget in Figure~\ref{fig:mf.versus.sigma}, we include 
the difference in number densities estimated by \citet{fontana:highz.mfs}, 
\citet{vandokkum06:drgs}, and 
\citet{marchesini:highz.stellar.mfs}, 
and variation in the distribution of profile shapes/sizes fitted in other 
works, including \citet{trujillo:ell.size.evol.update,toft:z2.sizes.vs.sfr,
buitrago:highz.size.evol,cimatti:highz.compact.gal.smgs}. 
These yield similar conclusions to within a 
factor $\sim2-3$. At the masses of interest, 
the observations should be reasonably complete to these 
high surface densities.

\begin{figure}
    \plotone{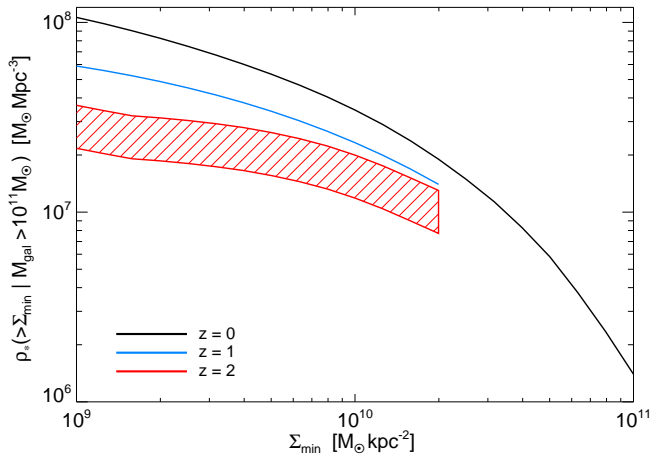}
    \caption{The total 
    global mass density of stars which reside in massive galaxies ({\em total} 
    galaxy mass $>10^{11}\,\msun$) and are above a threshold 
    stellar mass surface density (within the galaxy) of $\Sigma_{\rm min}$. 
    This is the integral of Figure~\ref{fig:mf.versus.sigma}, including only massive 
    galaxies. 
    We show the results from observations at 
    $z=0$ (black), $z=1$ (blue) and $z=2$ (red; shaded 
    range corresponds to typical uncertainties in the total mass density of 
    these compact high-$z$ systems). The high-$z$ 
    data is limited to the highest $\Sigma$ directly observed (limited by 
    resolution and seeing). 
    At $z=0$, $\sim4\times10^{7}\,\msun\,{\rm Mpc^{-3}}$ is locked 
    in massive ellipticals above a surface density of $>10^{10}\,\msun\,{\rm kpc^{-2}}$. 
    This is a factor of a few larger than the mass density at such 
    $\Sigma$ at $z\sim2$. 
    \label{fig:rho.versus.sigma}}
\end{figure}

We can also integrate the mass functions in 
Figure~\ref{fig:mf.versus.sigma} to obtain the total 
volume density of stellar mass in spheroids above 
some threshold in surface density $\Sigma_{\rm min}$; 
this is shown in Figure~\ref{fig:rho.versus.sigma}. 
Whereas mergers 
will not conserve number density, they should 
conserve total stellar mass in this calculation (to the extent that they 
do not change the $\Sigma_{\rm min}$ of the central regions of galaxies). Since 
the high-redshift systems are primarily massive, $>10^{11}\,\msun$, 
and their descendants cannot presumably be much lower 
mass, we restrict this calculation to only systems 
with a {\em total} stellar mass above this limit (although this only 
removes the very lowest-mass contributions to the high-$\Sigma$ 
population in Figure~\ref{fig:mf.versus.sigma}, and does 
not substantially affect our comparison). We then calculate, 
in systems above this mass, the total stellar 
mass above each threshold $\Sigma_{\rm min}$. 

Figure~\ref{fig:rho.versus.sigma} shows that there 
is $\sim4\times10^{7}\,\msun\,{\rm Mpc^{-3}}$ 
of stellar mass in $>10^{11}\,\msun$ ellipticals, above a surface density 
threshold of $\Sigma>10^{10}\,\msun\,{\rm kpc^{-2}}$ in the local 
Universe. This is comparable to the {\em total} stellar mass 
density of high-redshift red spheroids \citep{labbe05:drgs,
vandokkum06:drgs,grazian:drg.comparisons,abraham:red.mass.density}. 
Convolving over the observed size distribution at $z\sim2$, 
the mass density above this threshold $\Sigma$ is of course lower 
$\sim1-2\times10^{7}\,\msun\,{\rm Mpc^{-3}}$. 
We obtain a similar result comparing to the $z=1$ observations 
from \citet{vanderwel:z1.compact.ell}, with the relevant constraints being 
at lower surface density as there is less relative evolution. 

Our comparisons indicate that the local Universe contains just as much, or
more, stellar mass at high surface densities as implied by observations of
high-redshift systems. It is thus possible that all of the high-stellar
mass density systems at high redshift can be incorporated into massive
ellipticals today, without any conflict with their observed number
densities or surface brightness profiles.  In fact, a reasonable amount of
high-surface density material must continue to be added to the elliptical
population, perhaps by gas-rich mergers, from high redshifts until
$z=0$. It is the subject of another study whether gas-rich mergers
produce an adequate amount of high-$\Sigma$ stellar mass in current
galaxy formation models to account for the observed growth of
$\rho_\star$ in Figure~\ref{fig:rho.versus.sigma}. 
Nonetheless, the high density material at high
redshifts is not inconsistent with the $z\sim 0$ data and therefore 
does not have to ``go away.''

\section{The Mass At Low Stellar Densities}
\label{sec:obs.fx}

Our results demonstrate that the difference between low and high-redshift 
spheroids does not arise in their central densities, but in the large 
envelopes of low surface brightness material observed in low-redshift systems. 
This is the origin of their larger effective radii. 

There are two natural ways of reconciling the low and high-redshift 
observations with the hypothesis that the high-redshift spheroids are 
the progenitors of today's ellipticals. First, the high-redshift systems 
may not have much low-density material at large radii; low-density material 
would then have to be accreted at lower 
redshifts via late-time mergers (minor or major) with gas-poor 
disks and ellipticals (i.e.\ lower-density systems). 
Such a scenario is feasible, and indeed expected -- if the initial 
spheroid-forming mergers are sufficiently gas-rich, 
there will be little low-density material 
from extended stellar disks to  
contribute to an extended envelope \citep{hopkins:cusps.ell}. And 
comparison of clustering 
properties, merger rates, and stellar populations 
all imply that these massive, high redshift systems {\em should} grow 
by a factor $\sim1.5-2$ via these channels between $z\sim2$ and $z=0$, 
more or less sufficient to account for the 
envelopes seen in Figure~\ref{fig:profiles.compare} 
\citep[see e.g.][]{vandokkum:dry.mergers,bell:dry.mergers,
zheng:hod.evolution,lin:mergers.by.type,conroy:hod.vs.z}. 

The second way of reconciling the low and high-redshift observations is that 
high-redshift systems do have material at low surface densities already, but it is 
not seen in present observations (in such a case, there 
would be much less than expected of a buildup of low-density material). 
In order to be as conservative as possible, 
in considering the possible degree of envelope buildup since high redshifts, 
we consider this possibility here. 

\begin{figure*}
    \plotside{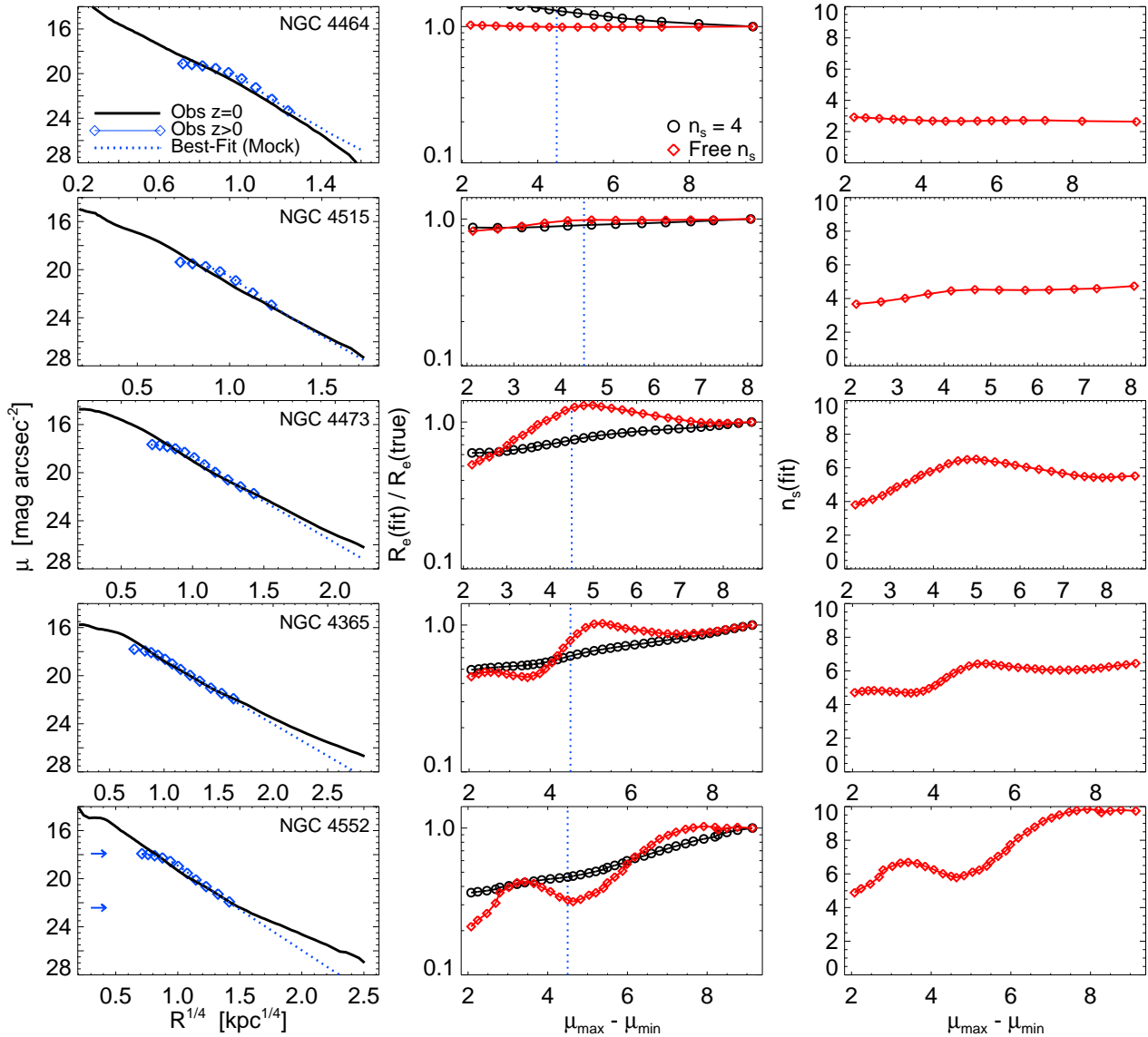}
    \caption{Best-fit light profile parameters $R_{e}$ and $n_{s}$ versus 
    image depth. 
    {\em Left:} Observed ($V$-band) surface brightness profiles (black) for several 
    typical local spheroids in the \citet{jk:profiles} 
    sample,$^{\ref{footnote:galaxydetails}}$
    from low mass (low-$n_{s}$, i.e.\ steep light profile decay; {\em top}) 
    to high mass (high-$n_{s}$, i.e.\ shallow decay/large envelope; {\em bottom}). 
    These are compared to the profiles constructed from mock images of the same 
    objects with best-case imaging available at $z\sim0.5-2$ 
    (cyan diamonds; seeing/PSF FWHM of $1.2$\,kpc, and depth/dynamic range of 
    $\mu_{\rm max}-\mu_{\rm min}=4.5$\,mag; arrows in {\em bottom left}). Dotted cyan line shows 
    the best-fit (convolved) Sersic profile fit to the mock high-$z$ imaging data, 
    extrapolated to large radii.     
    {\em Center:} Effective radii of the best-fit Sersic profile, 
    relative to the true $R_{e}$, as a function of image depth. 
    We construct a series of mock profiles with similar PSF, seeing, 
    noise, and pixel size (as {\em left}), but vary the image 
    depth $\Delta\mu\equiv\mu_{\rm max}-\mu_{\rm min}$.    
    We fit the resulting mock profile (with each $\Delta\mu$) with either a 
    free Sersic index $n_{s}$ (red diamonds) or fixed $n_{s}=4$ 
    (black circles). Dotted vertical line shows the $\Delta\mu$ of the mock 
    profile at {\em left}.
    {\em Right:} Corresponding best-fit $n_{s}$ (for free-$n_{s}$ fits). 
    At low masses (little stellar envelope), there is little dependence on the image depth. 
    At high masses (large-envelope systems) a strong trend appears, 
    because the $z=0$ profiles of massive 
    galaxies are fundamentally {\em not} well-described over 
    a large dynamic range by single Sersic laws. At small 
    $R\lesssim2-10\,$kpc ($\Delta\mu\sim2-4$), the radii that dominate the 
    fit in all but the deepest local imaging data, the profiles show 
    a steeper falloff (indicative of $n_{s}\lesssim4$, weak-envelope profiles) 
    and yield a smaller best-fit $R_{e}$. Only at 
    $R\gtrsim20-50\,$kpc ($\Delta\mu\sim6-8$) do the low-density wings 
    appear, leading to larger $n_{s}$ and $R_{e}$. 
    \label{fig:fit.vs.sb.limits}}
\end{figure*}

Figure~\ref{fig:fit.vs.sb.limits} illustrates 
circumstances under which the high-redshift observations may not be sensitive 
to extended, low-$\Sigma$ wings. We consider a 
few representative galaxy profiles from the local sample of 
\citet{jk:profiles}, ranging from low-mass cusp ellipticals with low Sersic indices 
at large radii (steep surface density falloff), to high-mass core ellipticals with high 
Sersic indices at large radii (extended envelopes).\footnote{\label{footnote:galaxydetails}
The specific five galaxies shown are (from top to bottom) 
NGC 4464, NGC 4515, NGC 4473, NGC 4365, and NGC 4552. 
The first three are classified as cusp ellipticals, the latter two 
as core ellipticals. The observations are described in 
\citet{jk:profiles}, but typically include $\sim80-100$ photometric 
points from a few pc to $\sim50$\,kpc in radii; photometric 
errors are $\lesssim0.04\,{\rm mag\,arcsec^{-2}}$ (not visible in 
Figure~\ref{fig:fit.vs.sb.limits}). 
They have stellar masses of $M_{\ast}=(0.17,\,0.13,\,1.2,\,3.9,\,2.0)\times10^{11}\,\msun$, 
true effective radii -- fit from the full data with proper multi-component 
profiles -- of $R_{e}=(0.60,\,1.05,\,3.19,\,14.6,\,10.6)\,{\rm kpc}$, and 
central velocity dispersions of 
$\sigma=(120,\,90,\,192,\,271,\,252)\,{\rm km\,s^{-1}}$. 
Fitting their {\em outer} profile shapes to 
Sersic profiles \citep[where they are uncontaminated by the central, high 
surface-density components; see e.g.][]{hopkins:cusps.ell,hopkins:cores}
yields best-fit outer Sersic indices of $n_{s}=(2.1,\,3.9,\,4.6,\,7.1,\,9.2)$.}
For each, we convolve the observed profile with a 
simple Gaussian PSF with typical best-case resolution for the high-redshift 
observations of interest ($1\,\sigma=0.5\,$kpc; FWHM $1.2$\,kpc). 
We then fit the (one dimensional) profile to a {\em single} Sersic 
profile (fitting appropriately convolved model 
profiles). We do this as a function of image depth. Specifically we 
plot the results of the fits as a function of $\Delta\mu\equiv\mu_{\rm max}-\mu_{\rm min}$, 
the surface brightness (${\rm mag\,arcsec^{-2}}$) of the 
deepest point included 
relative to the central/maximum surface brightness of the convolved image.
In Figure~\ref{fig:fit.vs.sb.limits} we plot the resulting 
best-fit effective radii, considering both fits with a 
free Sersic index $n_{s}$ and a fixed $n_{s}=4$. For fits 
with a free $n_{s}$, we plot the corresponding best-fit $n_{s}$.

For low-mass  and intermediate-mass galaxies, which have  
low $n_{s}$ and therefore fall off steeply in $\Sigma$ at large $R$, 
the true profiles can be recovered with even relatively shallow 
observations. If anything, for the lowest mass-galaxies, which 
have Sersic profile indices $n_{s}\sim2-3$ 
\citep[when fit in this manner to a single Sersic index; see e.g.][]{balcells:bulge.scaling,
balcells:bulge.xl,ferrarese:profiles}, 
the effects of the PSF tend to slightly {\em increase} the inferred effective 
radius in shallow images. For systems very close to 
$n_{s}=4$, characteristic of intermediate-mass galaxies, there is almost 
no depth-dependent bias to the inferred $R_{e}$ and $n_{s}$ 
(see NGC 4515 in Figure~\ref{fig:fit.vs.sb.limits}). 
The same conclusions are reached in the analysis of 
mock-redshifted SDSS images of low-mass galaxies 
in \citet{vanderwel:z1.compact.ell}. In short, if there is not a 
pronounced low-surface brightness envelope, then whether or not 
the low-surface brightness data is well-observed makes little difference 
to the fitted properties. Again, we emphasize, for {\em normal} 
$\sim L_{\ast}$ ellipticals, there appears to be no bias, and the 
relevant parameters are recovered well via simple Sersic fits, even 
given a very limited dynamic range of fitting. 

However, in the most extreme systems characteristic of the 
most massive galaxies -- those with large 
outer Sersic indices indicative of extended envelopes that include 
significant low-density material out to $\sim100\,$kpc radii -- 
the inferred $R_{e}$ and $n_{s}$ can be sensitive to whether or 
not this material is included in the fits. 
Considering the entire 
sample, we find that the typical depth required to obtain a ``converged'' $R_{e}$ and 
$n_{s}$ (to within $\sim30-50\%$ of the value obtained with the 
deepest available data) is a strong function of stellar mass (really, the 
strength of the low-density envelope), 
rising from $\Delta\mu\sim4\,$mag in intermediate-mass systems 
to $\Delta\mu\sim6-8$ in high-mass ($M_{\ast}>10^{11}\,\msun$) 
systems. If we require that at least $50\%$ of local systems 
have converged $R_{e}$, we obtain the approximate 
mass-dependent criteria
$\Delta\mu_{\rm conv}>3.0+2.3\,\log{(M_{\ast}/10^{11}\,\msun)}$; 
if we raise our desired threshold to $\sim75-90\%$, 
the minimum $\Delta\mu$ should be uniformly deeper by another 
$1.5\,$mag. In terms of physical radii, depth of $\Delta\mu\sim4$\,mag
corresponds roughly to a maximum well-sampled, high $S/N$ 
physical radius of $\sim5$\,kpc; 
the depth required for obtaining converged $R_{e}$ in massive systems  
($6-8$\,mag) corresponds to physical radii $\gtrsim20-50$\,kpc.

This bias arises because the systems with large envelopes 
are {\em not} perfect Sersic profiles; indeed, it is now well-established 
that {\em no} populations of spheroids are completely described by 
single Sersic profiles given sufficient dynamic range 
\citep[see e.g.][]{kormendy99,
ferrarese:profiles,lauer:bimodal.profiles,cote:smooth.transition,jk:profiles}. 
As a consequence, although a single Sersic profile is often 
a formally good fit (in a $\chi^{2}$ sense) 
over limited dynamic range, the best-fit $n_{s}$ can change systematically 
as that dynamic range changes. The effect is most pronounced in 
systems with the most dramatic envelopes; 
at small radii $R\lesssim2-10\,$kpc, 
systems are either ``concave down'' (which corresponds to a local Sersic index 
$n_{s}<4$) or uncurved ($n_{s}=4$). At larger radii, the profiles become 
more ``concave up'' ($n_{s}>4$). 
It is important to stress that, as such, the ``correct'' Sersic index is fit 
for the dynamic range sampled in the simple examples shown. 
If, in fact, the high-mass systems shown in Figure~\ref{fig:fit.vs.sb.limits} were 
perfect Sersic profiles, there would be no significant effect. 
Ultimately, this reflects the obvious caveat to any limited dynamic-range 
observation: a fit is being extrapolated to radii not directly observed, 
based on some assumed functional form. 

We emphasize that Figure~\ref{fig:fit.vs.sb.limits} is intended to be purely 
illustrative; we are {\em not} attempting to construct a specific 
comparison with or calibration for any individual observed sample. 
The specific calibrations for these observations are different 
sample to sample, and well outside the 
scope of this paper -- we refer to the relevant observational papers 
for more details. For example, the real fit results will also depend on 
the resolution, instrument PSF, sky subtraction, and other details. 
Given the non-trivial dependence of profile shape curvature on 
radius in the most extreme systems, the results will depend 
just as much on the relative weighting (error bars as a function of radius) 
in the observed profile as on any choice of a ``cutoff'' radius (recall, we 
simply truncate the profile at some limit -- realistically, this will appear 
as some surface-brightness-dependent error bar). 
Moreover, in many works \citep[e.g.][]{trujillo:ell.size.evol.update,
vandokkum:z2.sizes,buitrago:highz.size.evol}, the best-fit Sersic profile 
is determined directly from a fit to the two-dimensional image data, 
assuming a (radius-independent) ellipticity. In other works 
\citep[e.g.][]{jk:profiles,lauer:bimodal.profiles}, 
the image data is used to produce a major-axis 
or circularized profile (which allows for e.g.\ variations in ellipticity with 
radius or isophotal twists) and then fit to a Sersic profile. 
The attendant systematics, at this level of detail, are not identical; 
and re-fitting the objects shown in Figure~\ref{fig:fit.vs.sb.limits} 
in two-dimensional images tends to suppress the run of Sersic index with 
fitted radius (owing to the difference in relative error weighting; 
C.\ Peng, private communication). 
The consequences of those details can, in principle, 
affect the inferred sizes in either direction, not just 
towards inferring smaller sizes in lower-depth observations. 

That being said, we can consider these caveats in the context of the 
deepest available observations at $z\sim1-2$, 
spanning $\Delta\mu\sim3-4\,$mag \citep[$S/N$ rapidly decreasing at 
$R\gg5$\,kpc; see][]{trujillo:size.mass.to.z3,cimatti:highz.compact.gal.smgs,
damjanov:red.nuggets}. 
At these depths, our comparisons suggest that -- if the 
``true'' profiles were identical to those of today's most extreme massive 
galaxies (again, systems with $\sim100\,$kpc envelopes and 
very large $n_{s}\sim8-10$ -- the effective radii could be 
under-estimated by factors up to $\sim2-3$. 
It is not, in principle, hard to recover fitted $R_{e}\sim2-3\,$kpc 
for such systems in this simplified experiment, for individual objects 
given limited dynamic range. Note, however, that the details are 
sensitive to the issues above: there is less of a discrepancy (in 
at least this idealized case), for example, 
comparing the two-dimensional profiles. 
More importantly, even allowing for this level of an effect, 
it is very difficult to account for the entire evolution observed; 
in short, there is a real deficit of low density material at large 
radii in the high-redshift systems. This is already apparent at 
the lowest $\Sigma$ sampled, in e.g.\ Figure~\ref{fig:profiles.compare}, 
where the high-redshift profiles are falling more rapidly than 
those of low-redshift analogues. 

Again, if there is no large envelope as in the extreme case 
considered here, then there will be no bias in $R_{e}$ 
in high-redshift observations. But of course, this becomes circular -- 
the point is simply that caution is warranted extrapolating any 
profile (especially profile shapes calibrated to the observations of 
low-redshift galaxies) to radii not directly observed. 

Other, simple tests can constrain these possibilities. For example, 
\citet{zirm:drg.sizes} stack $\sim14$ high-redshift ($z\gtrsim2$) systems identified as compact 
(gaining roughly an additional $\Delta\mu\sim1.5$), and see 
no significant change in Sersic profile or size (relative to the 
luminosity-weighted average size of the individual fits). Likewise, 
stacking the $z=1$ compact systems in \citet{vanderwel:z1.compact.ell}, 
the best-fit Sersic index is $n_{s}\approx4$, comparable to the 
individual fit results. These appear to support what can 
already be seen in Figure~\ref{fig:profiles.compare} 
\citep[and Figure~1 of ][]{bezanson:massive.gal.cores.evol}; at the maximum radii 
$\sim 5\,$kpc sampled in the high-redshift observations, the profiles 
do appear to be falling more rapidly with radius than profiles 
of similar-mass low-redshift galaxies. 

It appears, therefore, despite the caveats above, 
that there is a real deficit of material 
at low surface densities at large radii around massive, 
high-redshift galaxies. 

However, it is important emphasize two simple points this highlights 
with respect to the  
interpretation of observed surface density profiles. 
First, some caution is always warranted when the dynamic range is 
limited. Additional checks, such as stacking the observed images, 
calibrating against low-redshift samples, and testing for bias in 
surface brightness limits of the sample, are important (and 
indeed have been a critical component of many of the high-redshift 
studies considered here). Given the lack of strong 
constraints on the amount of material at very low surface densities 
at high redshifts, the effective radii of the most extreme systems 
could (again, depending on the circumstances as noted above) 
be biased at the factor $\sim2$ level. This is insufficient to explain 
the total observed evolution -- there appears to clearly be some real evolution 
in the amount of low surface density material -- but it points to the 
still relatively large uncertainties in just ``how much'' low density 
material is (or is not) already in place at $z=2$. Constraints 
on this quantity, in particular as a function of redshift, 
will be of considerable interest for constraining models 
of how low-density envelopes build up from high redshifts to today. 

Second, it should be borne in mind that real ellipticals 
are not always perfect Sersic profiles, and in some cases 
the best-fit profile shape can change depending on the dynamic 
range sampled. As such, direct comparison of profiles, such 
as that considered here or in e.g.\ 
\citet{bezanson:massive.gal.cores.evol} are important (as opposed to just comparison 
of fitted quantities such as Sersic index and effective radius). 
Moreover, inferences made from extrapolation of Sersic 
fits should always be treated with some caution: the approach 
appears to work reasonably well with normal, $\sim L_{\ast}$ galaxies, 
for inferring simple quantities such as the effective radius, but 
it ultimately depends on the {\em assumption} that the functional 
form of the Sersic profile is a good description of the ``true'' profile 
at radii where the direct observational constraints are not 
as strong. For example, the approach of many works 
attempting to infer whether or not significant bias is present, by 
mock imaging systems modeled as single Sersic profiles 
(artificially redshifting and imaging them, then re-fitting to a 
Sersic profile), is a useful exercise, but does 
not necessarily capture all of the physical possibilities. 
Even at low redshifts, in fact, only a small fraction of 
massive ellipticals have effective radii that can be {\em directly} 
determined from the observations \citep[i.e.\ converged $R_{e}$ 
from the actual observed light profile, independent of fitting or 
extrapolation, that do not change as deeper radii are sampled; 
see e.g.\ the discussion in][]{jk:profiles}. Extending these samples, 
in particular at low and intermediate redshifts, is important both 
for studying the low-density envelopes that build up at later 
times and for informing our interpretation of the 
high-redshift observations.

\section{Discussion}
\label{sec:discussion}

\subsection{Can Yesterday's Compact Spheroids Be the Cores of Today's Ellipticals?}
\label{sec:discussion:compactness}

We have compared the stellar mass profiles of high-redshift, ``compact'' massive 
spheroids and well-studied local massive ellipticals. 
There has been considerable debate in the literature regarding the origins and 
fate of the compact high-redshift systems: they appear to be smaller ($R_{e}\sim1\,$kpc 
at $\sim10^{11}\,\msun$) than all but a tiny fraction of local, similarly 
massive galaxies which have $R_{e}\sim4-5\,$kpc at $z=0$. 
However, comparing the surface stellar mass density profiles directly, over the 
ranges that are actually well-sampled by observations, we show that the 
observed high-redshift systems have surface density profiles similar to the inner, 
dense regions of local massive ellipticals (Figure~\ref{fig:profiles.compare}). 

In other words, although the {\em effective} 
stellar mass surface density within $R_{e}$, 
$\Sigma_{\rm eff}\equiv1/2\,M_{\ast}/(\pi\,R_{e}^{2})$, is large 
in high-redshift systems, the 
physical stellar surface densities are 
comparable to the typical central surface densities observed at 
radii $\sim0.5-5$\,kpc in many local ellipticals. 
The centers or cores of local spheroids are just as dense as 
those of high-redshift systems: 
the difference is in the effective radii and effective densities, owing to the 
large, extended wings/envelopes of low surface brightness 
material around local massive spheroids. This material leads to a 
larger $R_{e}$ and lower effective surface density in the 
local systems. 

We have further shown that the distributions of the {\em maximum}
stellar surface densities are nearly the same at $z\sim2$ and $z=0$
(Figure~\ref{fig:peak.densities}) -- at small radii today's ellipticals
have similar maximum nuclear stellar surface densities of
$\sim0.3-1\times10^{11}\,\msun\,{\rm kpc^{-2}}$ over a wide range in
stellar mass from $\sim10^{9}\,\msun$ to $\gtrsim10^{12}\,\msun$.  The high
redshift systems have their central surface densities smeared out by PSF
and seeing effects, and thus do not reach these densities at any observed
point; but extrapolating their best-fit Sersic profiles inwards they
exhibit similar peak surface densities. 
Similar conclusions are 
reached by \citet{bezanson:massive.gal.cores.evol} as well (note that the factor 
$\sim$a couple apparently higher densities those authors note 
in the central regions of high-redshift systems depends on extrapolating 
Sersic profiles to smaller radii than observed, as well as ignoring the 
scatter in the central densities of ellipticals today). 
High-redshift red galaxies are
thus not uniformly more dense; indeed, the maximum/peak surface density of
spheroids does not appear to evolve significantly from $z\gtrsim2$ to
$z\sim0$. 

Using a large sample of local, high-dynamic range observations, 
we have constructed a census of the local spheroid population and 
have quantitatively calculated the number of systems with central/core 
mass densities above a given 
surface mass density and stellar mass threshold 
(Figures~\ref{fig:frac.above.sigma}-\ref{fig:rho.versus.sigma}). 
We have used this to construct the stellar mass function of 
spheroid ``cores'' -- i.e.\ the stellar mass function of the 
parts of today's ellipticals that lie above a given surface stellar mass 
density threshold. 

The regime of particular interest is $\Sigma\sim10^{10}\,\msun\,{\rm kpc^{-2}}$, 
which corresponds to the effective surface brightness of the 
high-redshift compact systems ($10^{11}\,\msun$ 
with $R_{e}=1\,$kpc). We find that 
$\sim25-35\%$ of the stellar mass density in $z=0$ 
massive spheroids 
lies above this surface density. Typical ellipticals have 
cores containing $\sim1-5\times10^{10}\,\msun$ 
above this threshold. 
Comparing this to the observed properties of massive galaxies at $z=1$ and $z=2$, 
we find that by both number and total stellar mass, 
all of the high-redshift, compact systems can be accounted for 
in the cores of today's ellipticals. For example, even in the extreme 
case in which every $z=2$, $10^{11}\,\msun$ or larger spheroid 
(space density $\approx10^{-4}\,{\rm Mpc^{-3}}$) 
had $R_{e}=1\,$kpc, this would 
correspond to the same space density of systems with 
$>1/2\,M_{\ast}=5\times10^{10}\,\msun$ above 
the effective surface density $10^{10}\,\msun\,{\rm kpc^{-2}}$. 
At $z=0$, the space density of such massive, high surface 
density cores is a factor $\sim1.5-2$ higher. 
Doing the calculation more properly (convolving over the 
mass function and distribution of profile shapes), there is a factor 
$\sim2$ more mass in local massive cores than is present at 
$z>2$; the difference is qualitatively similar, but 
smaller, comparing to $z=1$ populations. Not only can the 
high-redshift systems be accommodated (rather than being 
destroyed), but high density material 
in the centers of ellipticals may continue to build up even at 
relatively low redshifts.

\subsection{How Does this Relate to Physical Models?}
\label{sec:discussion:models}

These conclusions are of considerable importance for 
physical models of spheroid formation and evolution, 
and in particular for the models that have been 
proposed to explain both the formation of high-redshift, 
apparently compact galaxies and their evolution into local 
$z=0$ systems. 

Models for spheroid formation naturally predict that 
ellipticals and bulges are fundamentally two-component 
objects, with a dense, central core built by dissipational processes -- 
the loss of angular momentum in a progenitor gas disk, which then 
falls to the center and turns into stars in a compact starburst -- 
and an extended, lower-density envelope build by 
dissipationless processes -- the violent relaxation of 
progenitor disk stars, observed to be at much 
lower phase-space densities than the compact cores of 
ellipticals \citep{mihos:cusps,hopkins:cusps.fp}. 
Observations in the local Universe have confirmed much of this 
picture and made it increasingly robust 
\citep{kormendy99,
hibbard.yun:excess.light,rj:profiles,jk:profiles}. Indeed, simulations 
by several independent groups 
consistently find that it is not possible to make realistic 
ellipticals without the appropriate mix of these two 
components \citep{barneshernquist96,naab:gas,
cox:kinematics,onorbe:diss.fp.details,
jesseit:kinematics,burkert:anisotropy}. 
As such, the existence of dense cores in spheroids at both low and 
high redshifts is a natural consequence of dissipational spheroid formation. 

It is possible, if mergers 
are sufficiently gas-rich\footnote{We 
exclude cases where the systems are gas-dominated from 
this discussion, as in these cases the relevant 
physics leads to qualitatively different behavior 
in mergers, and will not necessarily make 
spheroids at all \citep{robertson:disk.formation,
hopkins:disk.survival,hopkins:disk.survival.cosmo}.}, to build just the 
high-density core, and to add the envelope 
in relatively gas-poor mergers at later times. 
Given the gas-richness of high-redshift galaxies, some size evolution is 
naturally predicted in models, with high-redshift 
systems being more dominated by the dense, 
dissipational remnant \citep{khochfar:size.evolution.model,hopkins:cusps.evol}. 

The high-redshift observations represent an ideal opportunity to 
catch such cores ``in formation'' and strongly constrain 
their physical origin. Today, such cores 
are typically extremely old: $\sim10\,$Gyr. 
At high redshifts, however, they have ages $\lesssim500\,$Myr 
\citep{kriek:drg.seds}. Understanding their stellar populations, 
metallicities, kinematics, and densities is critical to inform models 
of how dissipation builds the central regions of galaxies. 
There appears to be a natural link between the 
observed compact red galaxies and bright sub-millimeter galaxies, which 
are intense starbursts with consistent number densities (accounting 
for their short duty cycles) and 
physical sizes \citep{tacconi:smg.maximal.sb.sizes,younger:highz.smgs,
younger:smg.sizes,cimatti:highz.compact.gal.smgs}. 
This class of SMGs is widely believed to be the 
product of major mergers \citep{shapiro:highz.kinematics,
tacconi:smg.mgr.lifetime.to.quiescent}. Establishing 
further connections between these populations would not only enable new 
tests of the merger hypothesis, 
but would also rule out alternative models (e.g.\ monolithic 
collapse) for spheroid core formation. 

The maximal mass densities of spheroids at both low and 
high redshift ($\sim10^{11}\,\msun\,{\rm kpc^{-2}}$), for example, may 
inform models of star formation and feedback in 
extreme environments. This maximum surface density is intriguingly similar 
to previous suggestions of maximal (Eddington-limited) 
starbursts: if the observed mass surface density is initially pure gas, forming 
stars according to the \citet{kennicutt98} relation 
(giving $\sim2500\,\msun\,{\rm yr^{-1}}\,{\rm kpc^{-2}}$),  
this implies a luminosity $=1.5\times10^{13}\,L_{\sun}\,{\rm kpc^{-2}}$. 
This is the Eddington limit for dusty systems 
\citep{thompson:rad.pressure}. And, interestingly, it corresponds 
to the maximum SFR surface density in dense SMGs 
\citep{tacconi:smg.maximal.sb.sizes,younger:smg.sizes}. 
The fact that so few ellipticals scatter above 
this peak surface density also suggests that their centers may have 
formed in a few dissipational events -- 
if Eddington-limited arguments explain these peak densities, 
then there is no reason why they could not be exceeded if the 
gas ``trickled in'' at a lower rate or in several smaller events. 
Clearly, it is of interest to investigate constraints on 
star formation and feedback models stemming from this. 

Understanding the evolution in profile shapes, in particular how 
central versus outer densities evolve, is necessary to constrain 
how the potential and binding energy at the centers of spheroids 
evolve. This may be intimately related to the BH-host galaxy correlations 
in feedback-regulated models of BH growth \citep[see e.g.][and references 
therein]{hopkins:bhfp.obs,hopkins:bhfp.theory}. 
There has been considerable debate regarding the 
state of the BH-host correlations at these redshifts: better understanding of 
spheroid cores that dictate the local potential depth is 
critical to inform theoretical models. 

Current observations suggest that a large fraction of 
the high-density material in spheroids was assembled at early times. 
Although we have shown that it is possible to accommodate the mass in 
dense, high redshift cores in the elliptical population today, the observations 
suggest that of order half the massive cores of 
today's massive ellipticals had to be in place by $z>2$. Compare 
this to just $\sim5\%$ of the total spheroid mass density in 
place at these redshifts, and $\sim20\%$ of the massive 
galaxy ($>10^{11}\,\msun$) density \citep{grazian:drg.comparisons,
perezgonzalez:mf.compilation,
marchesini:highz.stellar.mfs}. In other words, it appears 
that the massive cores of today's ellipticals assembled preferentially 
early. This is qualitatively consistent with models of dissipational formation, 
but in semi-analytic models explaining 
early massive elliptical formation is quantitatively quite challenging 
\citep[see e.g.][]{bower:sam,fontana:highz.mfs}. 
The observations thus constrain not just the total 
assembly, but how this takes place -- invoking early minor mergers or 
gas-poor processes, for example, might be able to 
account for the shape of the mass function, but 
would not explain the early formation of dense cores. 

The observational comparison here favors models in which  
high-redshift compact galaxies 
are not destroyed (as has typically 
been concluded), but accrete or reveal previously ``hidden'' extended 
envelopes of low surface-brightness material. Their central densities remain, but 
with the appearance of low-density material, the effective radii and 
profile shapes quickly become comparable to massive galaxies today. 
If the high redshift systems genuinely do not have such low-density envelopes, 
the required evolution is only a factor $\sim1.5-2$ growth in stellar mass 
from high redshifts to today. 
Indeed, dry mergers onto such massive, 
early-forming systems are cosmologically inevitable, and the 
mass growth requirements found here are consistent with 
the current stellar mass function constraints 
\citep{brown:hod.evol,perezgonzalez:hod.ell.evol}. 
Moreover most dry mergers in such massive systems 
will be with later-forming, less massive and less dense systems, that 
are not expected to disrupt the dense cores, but will instead build up 
an envelope of lower surface density material 
\citep[see e.g.][]{naab:size.evol.from.minor.mergers}. 

In fact, the observations allow the possibility that the 
central regions of some massive ellipticals today have evolved with little or no 
strong perturbation, and corresponding change 
in densities, since high redshifts, while their envelopes assemble 
over this same period in time.  
This is important for the kinds of mergers that influence 
bulge formation: mergers with the kinds of low-density systems 
needed to build up an extended, low-density envelope 
(minor mergers, mergers with relatively gas-poor disks, and 
mergers with later-forming, less dense ellipticals) will not 
significantly perturb the (much more dense) central regions of the 
galaxy. In detail, even equal mass mergers with equal density systems 
do significantly less to lower than central densities than 
a simple energetic argument would imply: the central density profile 
remains relatively unperturbed while energy tends to be preferentially 
transferred to less bound outer material \citep[see][]{hopkins:cores}. 

Such mergers may also transform an initial central ``cuspy'' profile 
into a ``cored'' profile, via the scouring action of a binary BH-BH merger 
\citep[if the initial high-redshift systems are ``cuspy,'' expected if they have 
just formed in gas-rich mergers, then they must become cored by 
low-redshift to correspond to observed systems; see e.g.][]{faber:ell.centers,
graham:core.sersic,cote:smooth.transition,jk:profiles}. Both 
minor and major dry mergers are expected to be efficient mechanisms 
for scouring, even where the secondary is sufficiently low density so as 
not to perturb the density profile as discussed above 
\citep[see e.g.][]{milosavljevic:core.mass}. But the cusp-core 
distinction (in all but the most extreme systems) affects the mass profile 
only on very small scales (well below the scales observed in all but the 
nearest systems) and involves only a very small fraction of the 
galaxy mass ($\sim M_{\rm BH}$, or $\sim10^{-3}\,M_{\ast}$).

\subsection{Observational Tests and Future Prospects}
\label{sec:discussion:outlook}

Observed high-redshift systems require the 
growth of extended, low surface density envelopes to match the 
profiles of massive ellipticals today. 
It is quite possible that such envelopes are entirely absent 
at high redshifts. 
However, Figure~\ref{fig:fit.vs.sb.limits} highlights 
the fact that it remains difficult to 
say precisely how much of a ``mass deficit'' at large radii 
must be accreted from $z=2$ to $z=0$. In short, the 
total mass at very low surface densities at high redshifts remains 
more uncertain than the mass budget at high surface densities. 
And even where observations 
detect light at relatively low surface brightness, the 
relevant $S/N$ weighting means that the 
profile fits can be preferentially weighted towards the 
high-brightness central regions. 
As a result, there could remain 
some non-trivial differences in the best-fit $R_{e}$ from 
high-redshift observations, compared to deep low-redshift observations 
of analogous systems. 

At low and intermediate stellar masses, where Sersic 
indices tend to be low ($n_{s}\lesssim4$), 
our simple experiments, as well as those in other works 
calibrated to the specific observation techniques used therein
\citep[see e.g.][]{zirm:drg.sizes,
trujillo:size.evolution,trujillo:ell.size.evol.update,
vanderwel:z1.compact.ell,cimatti:highz.compact.gal.smgs}, 
suggest that there is no strong dependence on 
the depth of the observations. In other words, 
when there is no significant envelope, 
the precise mass profile of the low-density material makes little difference. 
At the highest masses, however -- i.e.\ local systems 
where the outer Sersic indices are very large ($n_{s}\gtrsim6$) 
because of the presence of large stellar envelopes that can 
extend to $\gtrsim100\,$kpc scales -- 
there can be a non-trivial mass at extremely low brightness, 
whose profile is difficult to recover in detail. 
If massive-envelope ellipticals such as NGC 4552 were present at 
$z>0.5$, full recovery of their envelopes (in 
the sense of recovering high S/N profile information) 
would require observations 
$\sim4-5$\,mag deeper than typical of individual 
high-redshift objects (Figure~\ref{fig:fit.vs.sb.limits}). 
Even at low redshifts, only a small fraction of objects 
at the highest masses have {\em directly} measured and converged 
effective radii (determined purely from the observed light profile); 
in other cases the effective radius is usually estimated by 
extrapolation of a best-fit Sersic profile to low densities 
and large radii. It is possible, if envelopes are present, but 
do not follow a perfect Sersic profile (or if the 
Sersic profile shape of the galaxy profile changes from 
small to large radii), to under-estimate the 
envelope mass, and correspondingly 
to under-estimate the true effective radius by a factor $\sim2$. 

This has important consequences even at low redshifts. Some 
recent studies have claimed that {\em most} of the apparent size-mass 
evolution in massive galaxies occurs at very low redshifts, 
a factor of $\sim2-3$ change in sizes 
between $z=0.1-0.3$, with the evolution 
from $z=0.4-2$ restricted to a smaller factor $\sim1.5-2$ 
\citep{bernardi:bcg.size.evol,ferreras:lowz.ell.size.evol}. Such extreme apparent low-redshift 
evolution (as opposed to the more plausible high-redshift 
evolution) may be related to the observed dynamic range 
and fitting: it is almost impossible to explain 
such a large change in the true stellar 
half-mass radius in any model over a narrow low-redshift interval $z=0.1-0.3$. 
Independent constraints have clearly established that in this 
redshift range, there is almost no evolution in the 
stellar mass function of massive spheroids, 
nor is there significant evolution in the (uniformly old) stellar 
populations, or any significant stellar mass loss or new 
star formation given the old stellar population ages 
\citep{thomas05:ages,nelan05:ages,
gallazzi06:ages,masjedi:merger.rates,borch:mfs,jones:lrgs}. Likewise, there is no change in 
their kinematics or the fundamental plane relation 
in clusters or the field \citep{treu:fp.evolution,alighieri:fp.evolution,
vanderwel:fp.evolution,vandokkum:fp.evol}. Even the 
dark matter halos grow by only a tiny 
fraction over the redshift range of $z=0.1$ to $0.3$ 
($<0.1$\,dex; almost all accreted at large radii). What 
do change over this redshift range, however, are the spatial resolution 
and effective surface brightness limits, and as a consequence 
the radial range of the profile sampled in observations. 
We find that the observations of rapid changes in the best-fit $R_{e}$ at low redshifts 
can be accounted for by the biases summarized in Figure~\ref{fig:fit.vs.sb.limits}. 
And indeed, other observations, using somewhat different 
fitting methodology and sample selection, have found no such significant evolution 
at low redshifts, while the high-redshift evolution appears more 
robust \citep[see e.g.][]{mcintosh:size.evolution,
trujillo:ell.size.evol.update,franx:size.evol}.

Clearly, better constraints on just how much low-surface brightness material 
is present, as a function of redshift, will improve our understanding of 
elliptical galaxy formation. 
Whether there has been a great deal of evolution in the amount of 
low-density material, or relatively little, either result is of considerable interest. 
If high-redshift systems have essentially no stellar mass at 
low densities ($\Sigma\lesssim 10^{9-10}\,\msun\,{\rm kpc^{-2}}$), 
then mergers forming spheroids must initially be very gas-rich, 
and mergers with lower-mass galaxies or evolved disks at high redshift 
must be relatively rare. Moreover, this would determine how much 
material would have to be built up by subsequent mergers, 
putting requirements on models for tidal destruction and 
minor mergers as well as later dry major mergers. 
If some envelopes are present at high redshifts -- perhaps not as 
much as at low redshifts, but still non-negligible in mass -- 
then it would strongly limit the amount of 
growth, accretion, and dry merging such systems could 
undergo at lower redshifts. 
To the extent that envelopes exist that contain significant mass 
not previously detected, it would increase the stellar masses of these 
galaxies, perhaps yielding tension with constraints from e.g.\ the 
galaxy mass function, and certainly presenting a challenge for models, 
which have difficulty explaining very early mass assembly without 
much subsequent growth. 

There are a number of observational means to study these possibilities. 
Most directly, observations can probe whether such material exists 
at high redshifts. 
In addition to profile fitting, which necessarily 
relies on the assumption that some functional form will represent a 
good approximation to light at radii where the profile shape cannot be 
strongly constrained directly, it should be possible to estimate 
that total light contribution from comparison of deep integrations of the 
total light/stellar mass in fixed physical annuli. 
Also, stacking the observed high-redshift systems, allowing for a 
gain of $\sim1-1.5$\,mag in surface brightness depth, seems to 
show a continued steep falloff in surface density ($n_{s}\lesssim4$), 
indicating that there is not a large undetected mass 
in envelopes at high redshift \citep{zirm:drg.sizes,vanderwel:z1.compact.ell}. 
This reinforces the conclusion that the envelopes must be predominantly 
built up at lower redshifts; 
unfortunately, it remains difficult to say precisely how much material is 
at low densities (other than to say that it is not a large fraction of the galaxy 
mass and cannot change $R_{e}$ at the level of evolution seen), 
or what the exact mass profile of the low-density material is 
(e.g.\ whether it corresponds to the inference from extrapolating the fitted 
Sersic profiles, cuts off more steeply, or obeys a power law-like distribution 
analogous to observed intracluster/halo light in low-redshift systems). 

Observations of major and minor dry merger rates offer complimentary 
constraints on how much material is added to these systems between 
high redshifts and today. In the local universe, studying the properties of these 
envelopes can inform models of their formation histories. For example, 
stellar population gradients and kinematics might reflect a more dramatic transition in 
properties if the envelopes form by late accretion onto earlier-forming cores. 
If the envelopes form early, they will be metal poor and 
have different $\alpha$-element abundances compared to late-accreting 
disk/outer bulge material. Some efforts have been made along these lines 
\citep[see e.g.][]{sanchezblazquez:ssp.gradients}, 
and gradients in e.g.\ $\alpha$-element abundances show significant diversity, 
suggesting that in some systems, envelopes formed quickly (perhaps in e.g.\ later-forming 
ellipticals, where a large envelope can be formed from relatively gas-poor disks), 
while in others, envelopes are contributed by systems with more extended 
star formation (e.g.\ late mergers with lower-mass or later-forming systems, as 
expected for systems described here). However, samples remain small, and the 
differences are subtle; larger samples are needed to correlate the behavior of 
these properties with other aspects of galaxies (e.g.\ galaxy mass, kinematics, 
cusp/core status, environment) that might provide an indication of whether or not they 
are descendants of similar high-redshift systems. Moreover, typical observations 
of these quantities extend only to relatively small radii ($\sim R_{e}$); it remains 
difficult to probe stellar populations in the low-density outer regions that 
constitute spheroid envelopes. 
Together, however, these observations offer promising avenues towards constraining 
these observationally challenging, low-surface brightness components. 

Corroborating evidence for the picture presented here 
has recently been presented in 
\citet{cenarro:sigma.of.highz.compact.ell}. 
The authors measure the average velocity 
dispersions of compact spheroids at $z=0-2$, and show that 
it evolves weakly relative to e.g.\ the naive expectation 
of models with uniform contraction/expansion 
(that $\sigma^{2}\propto M_{\ast}/R_{e}$). Despite factor $\sim6$ evolution 
in effective radii of massive galaxies over this interval, 
the median $\sigma$ at fixed mass evolves by a factor $\approx1.3$. Moreover, 
the high-redshift $\sigma$ values are in fact 
consistent with the high end of the observed low-redshift scatter at this 
mass, or with the median $\sigma$ of slightly more massive 
(factor $\sim2$) low-redshift ellipticals. 
This is precisely the behavior expected if the central, high-density regions 
(which determine the central potential depth and correspondingly $\sigma$)
assemble preferentially early, and evolve relatively unperturbed 
to low redshift \citep[for examples in simulations, see][]{hopkins:cusps.evol}. 
Low-density material accreted later is clearly important for the effective 
radii, and changes e.g.\ the dark matter fraction within $R_{e}$ 
as well as the profile shape, but has very little effect on 
the velocity dispersion. 
The observations represent a strong constraint on 
the physical densities at fixed radius: 
if the central densities at high redshift were in fact substantially  
higher than those at low redshift, by even a factor of a few, 
it would significantly over-predict the observed evolution in 
velocity dispersions. 

Throughout this paper, we have neglected several additional physical effects 
that might affect estimates of the stellar surface mass densities and 
effective radii of high-redshift ellipticals. For example, at low redshifts, 
stellar mass-to-light ratios are independent of radius to within $\sim20\%$ in 
optical bands in massive ellipticals, which tend to be uniformly 
old and have weak color gradients \citep[see e.g.][]{faber:catalogue,
bender:ell.kinematics.a4}. However, given the observed stellar population gradients 
run backwards in time, or considering local recent merger remnants 
and/or simulations, the expectation is that 
this could be quite different at the high redshifts 
when the ellipticals formed \citep{schweizer96,rj:profiles,
yamauchi:ea.gradients}. The young, post-starburst 
population (ages $\lesssim1-2$\,Gyr) in the core 
has higher optical $L/M_{\ast}$, which can lead to a rest-frame $B$-band 
best-fit $R_{e}$ that is a factor $\sim1.5-2$ smaller than the 
stellar-mass $R_{e}$ (or the $B$-band $R_{e}$ observed when the 
system has aged and this effect vanishes). 
Some reassurance, however, comes from the fact that 
the size-mass relation does not seem to depend strongly 
on the precise stellar population 
age/colors \citep[specifically 
within the ``quenched'' or ``star-forming'' populations; 
see][]{perezgonzalez:hod.ell.evol,
toft:z2.sizes.vs.sfr} and the fact that the observed sizes of the 
galaxies of interest here are similar in restframe near-UV and optical 
\citep{trujillo:ell.size.evol.update}. 
There may also be some bias in 
estimated stellar masses of systems at similar ages, owing 
to the uncertain contribution of AGB stars 
\citep{maraston:ssps,maraston:ssp.effects}. This is estimated to 
be a possible factor $\sim1.2-1.4$ over-estimate of the high-redshift 
masses where high-quality optical photometry is available 
\citep{wuyts:irac.drg.colors,wuyts:photometry.biases.mgrrem}. 
We have conservatively ignored these effects, 
but if present they will strengthen most of our conclusions. 
To test for these effects, 
it is important to obtain deeper spectra and photometry, to 
test for the presence of blue cores and constrain the contribution of 
stellar populations of different ages (and stellar population gradients) 
in systems at intermediate and high redshifts.

\acknowledgments 
We thank Todd Thompson, Ignacio Trujillo, 
Chien Peng, Pieter van Dokkum, 
and Marijn Franx for helpful discussions. We also 
thank the anonymous referee for comments and suggestions 
regarding the treatment of observational effects. 
Support for PFH was provided by the Miller Institute for Basic Research 
in Science, University of California Berkeley. 
EQ is supported in part by NASA grant NNG06GI68G and 
the David and Lucile Packard Foundation.
\\

\bibliography{/Users/phopkins/Documents/lars_galaxies/papers/ms}

\end{document}